\documentclass{emulateapj}
\usepackage{natbib}
\usepackage{graphicx}
\usepackage{subfigure}
\usepackage{multirow}
\defcitealias{Brown98}{B98}
\usepackage[usenames]{color}
\usepackage{amssymb}

\newcommand{\Teff}{\mbox{$T_{\rm eff}$}}

\newcommand{\um}{\mbox{$\mu$m}}

\slugcomment{Version: \today}

\shorttitle{A Lack of C Stars in M31}
\shortauthors{Boyer et al.}

\begin{document}


\title{Is there a Metallicity Ceiling to Form Carbon Stars? -- A novel
  technique reveals a scarcity of C-stars in the inner M31
  disk\footnotemark[$\star$]}



\author{
M.~L. Boyer\altaffilmark{1,2,$\dagger$},
L. Girardi\altaffilmark{3}, 
P. Marigo\altaffilmark{4}
B.~F. Williams\altaffilmark{5},
B. Aringer\altaffilmark{6},
W. Nowotny\altaffilmark{6},
P. Rosenfield\altaffilmark{5},
C.~E. Dorman\altaffilmark{7},
P. Guhathakurta\altaffilmark{7},
J.~J. Dalcanton\altaffilmark{5},
J.~L. Melbourne\altaffilmark{8}, 
K.~A.~G. Olsen\altaffilmark{9},
D.~R. Weisz\altaffilmark{5}, 
}

\altaffiltext{1}{Observational Cosmology Lab, Code 665, NASA Goddard Space Flight Center, Greenbelt, MD 20771, USA}
\altaffiltext{2}{Oak Ridge Associated Universities (ORAU), Oak Ridge, TN 37831, USA}
\altaffiltext{3}{Osservatorio Astronomico di Padova -- INAF, Vicolo dell'Osservatorio 5, I-35122 Padova, Italy}
\altaffiltext{4}{Department of Physics and Astronomy G.\ Galilei, University of Padova, Vicolo dell'Osservatorio 3, I-35122 Padova, Italy}
\altaffiltext{5}{Department of Astronomy, University of Washington, Box 351580, Seattle, WA 98195, USA}
\altaffiltext{6}{University of Vienna, Department of Astrophysics, T\"urkenschanzstra{\ss}e 17, A-1180 Wien, Austria}
\altaffiltext{7}{University of California Observatories/Lick Observatory, University of California, 1156 High St., Santa Cruz, CA 95064}
\altaffiltext{8}{Caltech Optical Observatories, Division of Physics, Mathematics and Astronomy, Mail Stop 301-17, California Institute of Technology, Pasadena, CA 91125, USA}
\altaffiltext{9}{National Optical Astronomy Observatory, 950 North Cherry Avenue, Tucson, AZ 85719, USA}

\footnotetext[$\star$]{Based on observations made with the NASA/ESA
  Hubble Space Telescope, obtained from the Data Archive at the STScI,
  which is operated by the AURA, Inc., under NASA contract NAS5-26555.}
\footnotetext[$\dagger$]{martha.boyer@nasa.gov}

\begin{abstract}
We use medium-band near-infrared (NIR) {\it Hubble Space Telescope}
WFC3 photometry with model NIR spectra of Asymptotic Giant Branch
(AGB) stars to develop a new tool for efficiently distinguishing
carbon-rich (C-type) AGB stars from oxygen-rich (M-type) AGB stars in
galaxies at the edge of and outside the Local Group. We present the
results of a test of this method on a region of the inner disk of M31,
where we find a surprising lack of C stars, contrary to the findings
of previous C star searches in other regions of M31. We find only 1
candidate C star (plus up to 6 additional, less
  certain C stars candidates), resulting in an extremely low ratio
of C to M stars (${\rm C/M} = (3.3^{+20}_{-0.1})$$\times$$10^{-4}$)
that is 1--2 orders of magnitude lower than other C/M estimates in
M31. The low C/M ratio is likely due to the high metallicity in this
region which impedes stars from achieving ${\rm C/O}>1$ in their
atmospheres. These observations provide stringent constraints to
evolutionary models of metal-rich AGB stars and suggest that there is
a metallicity threshold above which M stars are unable to make the
transition to C stars, dramatically affecting AGB mass loss and dust
production and, consequently, the observed global properties of
metal-rich galaxies.
\end{abstract}


\keywords{stars: carbon --- stars: AGB and post-AGB --- stars: late-type --- galaxies: individual (M31)}


\section{Introduction}
\label{sec:intro}

Asymptotic Giant Branch (AGB) stars play a significant role in
galaxies' observed properties and their evolution. They are
responsible for a major share of galaxy luminosity
\citep[e.g.,][]{Maraston+06,Marigo+10,Boyer+11,Melbourne+12,MelbourneBoyer13}
and contribute considerably to the chemical enrichment of the
interstellar medium \citep[e.g.,][]{Marigo01,Ventura+01,KarakasLattanzio07}.
Despite its widespread importance, the AGB phase remains among the
most uncertain phases of stellar evolution modeling, leading to the
largest uncertainties in galaxy stellar population synthesis
\citep{Conroy+09}.


\begin{figure}
\includegraphics[width=0.98\columnwidth]{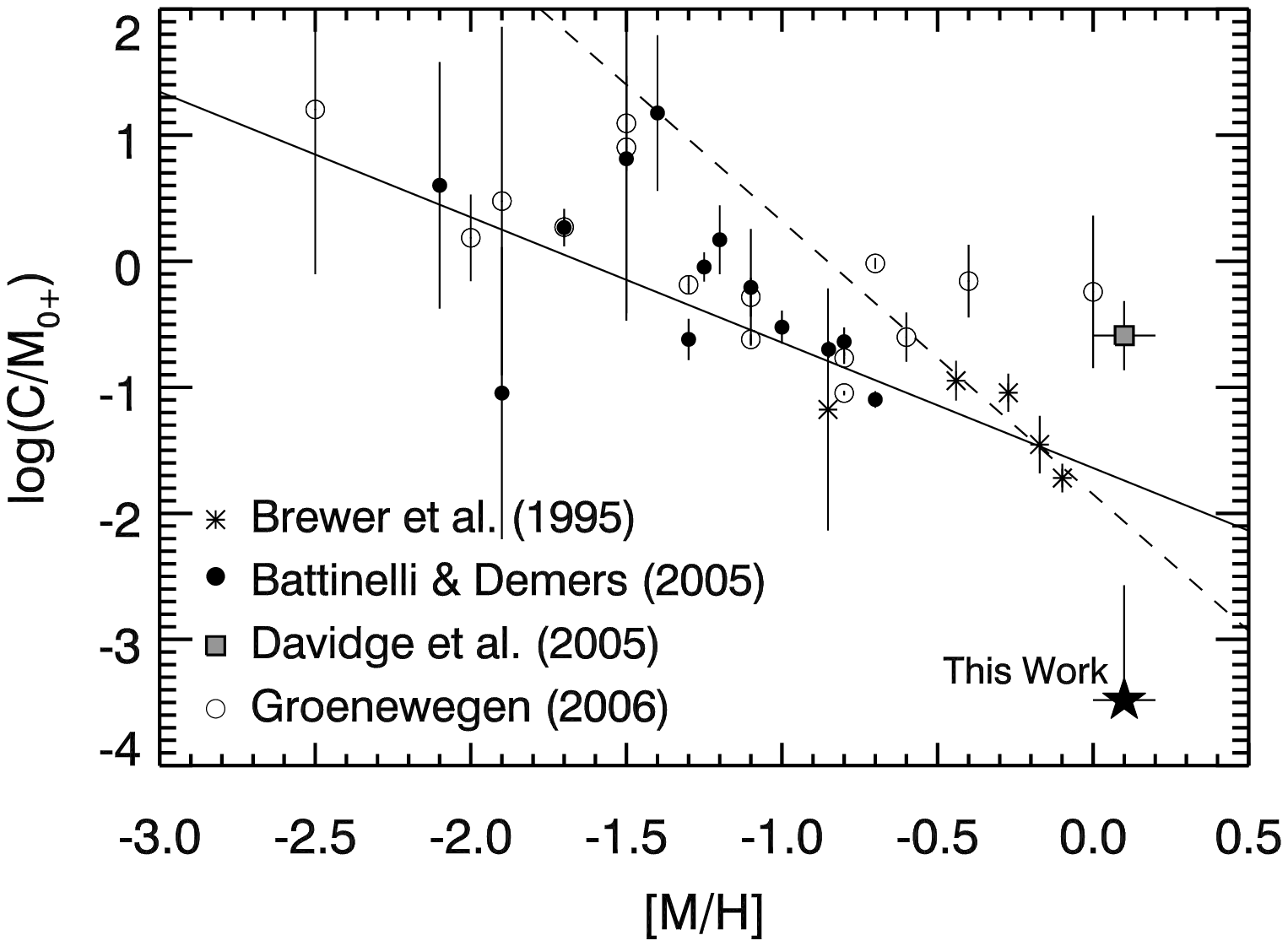}
\caption{The C/M ratio, adapted from \citet{Groenewegen06}
  and \citet{BattinelliDemers05} for nearby galaxies. Data for five
  regions in M31 from \citet{Brewer+95} are plotted with asterisks,
  with metallicities from \citet{ZuritaBresolin12}. These data
  were derived using optical photometry, so the number of C stars is
  underestimated by $\approx$20--60\% \citep[e.g.,][]{Boyer+09}. The
  error bars are derived solely from Poisson statistics and do not
  include information on detection biases or photometric
  uncertainties. The solid line is fit through all points, and the
  dashed line is fit only through the \citet{Brewer+95} points. The
  \citet{Davidge+05} point (gray square) and the point from this work
  (star) are derived from NIR photometry, with the metallicity
  range from \citet{Saglia+10}. We note that the metallicities for M31
  are not measured directly for the fields included here, but are
  instead derived from estimates of the metallicity gradient across
  the disk \citep{ZuritaBresolin12} or the bulge
  \citep{Saglia+10}.}
\label{fig:cm}
\end{figure}

\begin{figure*}
\includegraphics[width=1\textwidth]{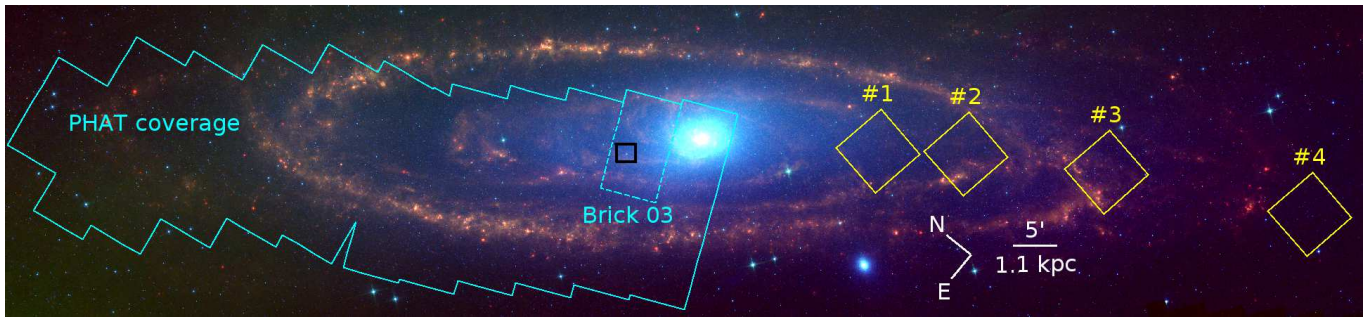}
\caption{3-color image of M31 from {\it Spitzer}
  \citep{Barmby+06}. 24~\um\ is plotted in red, 5.8~\um\ in green, and
  3.6~\um\ in blue. The PHAT coverage is shown in cyan
  \citep{PHATpaper}. The coverage discussed here is shown in black and
  is part of the PHAT survey's ``Brick 03''. Four of the regions from
  \citet{Brewer+95} plotted in Figure~\ref{fig:cm} are shown in
  yellow. Region \#5 is off the image to the southwest.}
\label{fig:img}
\end{figure*}

AGB atmospheres have complicated elemental abundances due primarily to
the third dredge-up (3DU) process, wherein they pull newly-synthesized
carbon and other elements to the surface. The carbon bonds with free
oxygen to make CO, and the excess oxygen or carbon dictates the
molecular and dust chemistry of the star
\citep{MarigoAringer09,FerrarottiGail06}, having drastic consequences
for the mass-loss and dust-production processes. The ratio (C/M) of C
stars (those AGB stars with excess free carbon in their atmospheres)
to M stars (those with excess free oxygen) generally decreases with
increasing metallicity \citep[e.g.,][]{Rowe+05,Groenewegen06,Groenewegen07,BattinelliDemers05}. Metal-poor
stars are more likely to become C-rich because less oxygen is
available to bind newly dredged-up carbon into CO and because the
depth of 3DU events increases at low metallicity
\citep{Karakas+02}. Both factors favor the formation of carbon stars
after fewer and fewer dredge-up events at decreasing
metallicity. Although this effect is well established on theoretical
grounds, quantitative predictions of the C/M ratio from first
principles are not straightforward, given the critical dependence of
C/M on the star formation history of the parent galaxy and the
uncertain details of 3DU modeling.  Empirically, the
relationship between the C/M ratio and metallicity is poorly
constrained owing to small number statistics and other sample biases,
especially at high metallicity (Fig.~\ref{fig:cm}).



\subsection{Carbon Star Surveys}
\label{sec:surveys}

Carbon star surveys in nearby dwarf galaxies have
helped inform evolutionary models at low metallicity \citep[e.g.,][and
  references therein;
    Fig.~\ref{fig:cm}]{Battinelli+07,BattinelliDemers09,Groenewegen06,Groenewegen07}. However,
the majority of these surveys were conducted entirely at optical
wavelengths ($<$9000~\AA) where dusty AGB stars are often undetected
owing to circumstellar extinction \citep[e.g.,][]{Nowotny+13}.  In the
Magellanic Clouds, $\approx$30\% of C stars are fainter than $M_{\rm
  I} = -2.5$~mag \citep{Boyer+11}, which is the limiting magnitude of
the surveys mentioned above. More complete AGB samples can be created
using observations in the near-infrared (NIR) where circumstellar dust
extinction is less severe. Several studies have searched
  for carbon stars using NIR photometry in metal-poor dwarf galaxies
  \citep[e.g.,][]{Cioni+06,Sibbons+12}. At higher metallicity, NIR
  searches are much more difficult owing either to the distance of the
  galaxy (in the case of M31) or to sample biases and uncertain
  distance measurements (in the case of the Milky Way).  

There are two examples of photometrically-complete NIR
  surveys of AGB stars in the inner region of M31:
  \citet{Stephens+03}, who used the {\it Hubble Space Telescope}
  NICMOS camera, and \citet{Davidge+05}, who used ground-based
  adaptive optics. Both of these surveys find examples of stars with
  very red $J-K$ colors. \citet{Stephens+03} identify these stars as
  long period variables, but do not comment on whether they are C or M
  stars.  \citet{Davidge+05} identify them as candidate C stars based
  on their NIR colors, suggesting the presence of a large C star
  population. This is in contrast to C star surveys in the Milky Way
  Bulge, which show an almost total lack of intrinsic C stars
  \citep[e.g.,][]{Azzopardi+91,Whitelock+93,Feast07,Miszalski+13}. Whether
  or not these red stars in the M31 bulge and inner disk are C stars
  remains unexplored, since narrow-band photometry and spectroscopy
  are difficult both due to limited sensitivity and crowding.



The WFC3 NIR imaging camera on board the {\it Hubble Space Telescope}
has allowed for high-resolution, sensitive imaging up to 1.6~\um,
enabling observations of AGB stars in galaxies farther than
$\sim$$5$~Mpc for testing and improving models of AGB mass loss
\citep[e.g.,][]{Melbourne+12}. However, it is impossible to
distinguish C- and M-type AGB stars with the available WFC3 broad-band
filters alone \citep{SNAPpaper,PHATpaper}. The medium-width filters
available for WFC3/IR, on the other hand, do provide the opportunity
to separate AGB subtypes (Sect.~\ref{sec:cstars}). NIR synthetic
spectra of AGB stars \citep[][and in preparation]{Aringer+09} suggest
that these filters can sample individual molecular features in C-rich
AGB stars. Therefore, these medium-band filters should provide an
efficient way of discriminating between C- and O-rich AGB stars.

To test this approach, we targeted a field from \citet{Davidge+05}
with the WFC3/IR medium-band filters so that any candidate C stars
detected could be compared to those identified via their $JHK$
colors. The \citet{Davidge+05} results suggest that a
  WFC3/IR single field should encompass a population of several
  hundred carbon stars, based on the fraction of AGB stars with red
  $J-K$ and $H-K$ colors in their small 22\arcsec\ field.  However,
our observations reveal the near-absence of C stars.

We have organized this paper as follows. In Section~\ref{sec:data}, we
detail our observations and analysis. Section~\ref{sec:discu}
describes our early results regarding the presence of C and M-type AGB
stars. Finally, the paper is summarized and the implications discussed
in Section~\ref{sec:conclu}.


\begin{figure*}
\includegraphics[width=\textwidth]{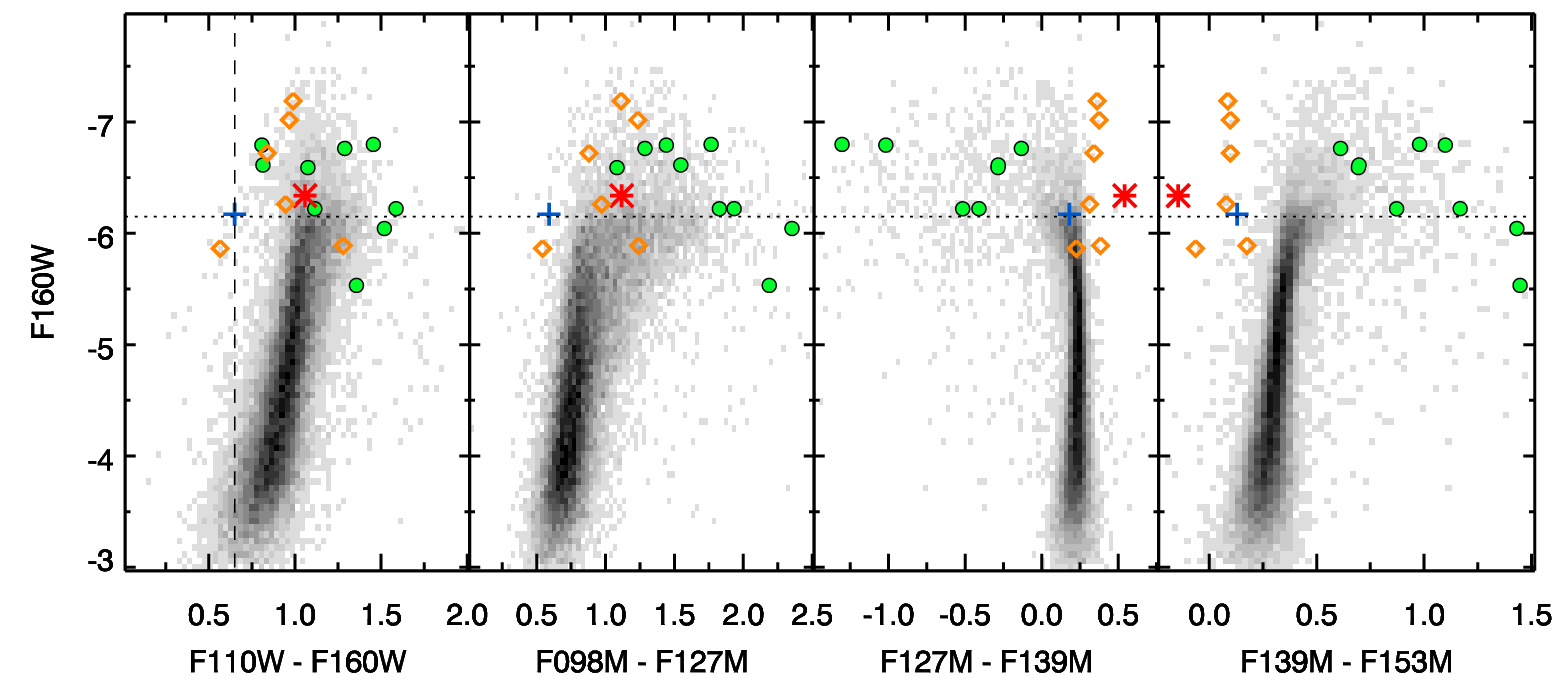}
\caption{Color-magnitude diagrams, plotted as Hess diagrams to show
  the point-source density. F160W and F110W magnitudes come from the
  PHAT survey \citep{PHATpaper}. Sources with $H-K>0.4$~mag from
  \citet{Davidge+05} are plotted as large green circles. The
  horizontal dotted line is the F160W TRGB from \citet{PHATpaper}. The
  one carbon star candidate identified in Figure~\ref{fig:ccdmodel} is
  shown as a red asterisk, and the 6 sources that fall on the edge of
  the region covered by the carbon star models in
  Figure~\ref{fig:ccdmodel} are plotted as orange diamonds. An
  additional star on the edge of M star model coverage in
  Figure~\ref{fig:ccdmodel} is plotted as a blue plus symbol. This
  star has colors consistent with a foreground M dwarf \citep[dashed
    line, also see Fig.21 from][]{PHATpaper}, as does one of the C
  star candidates.}
\label{fig:cmds}
\end{figure*}

\section{The Data}
\label{sec:data}

Data were collected in a Cycle~20 HST program (GO-12862, P.I.: Boyer)
and consist of 600~sec total integrations (300~sec $\times$ 2 frames)
in each one of the medium-band WFC3/IR filters: F098M, F127M, F139M,
and F153M. The 2\farcm3$\times$2\farcm1 field of view is centered at
$\alpha = 00^{\rm h}43^{\rm m}21\fs7$, $\delta =
+41\degr21\arcmin46\farcs2$, approximately 2~kpc from the center of
M31 (Fig.~\ref{fig:img}). Photometry was produced using the software
package DOLPHOT 2.0 \citep{Dolphin00}, which uses {\sc TinyTim}
point-spread functions \citep[PSFs;][]{Krist95,Hook+08} to measure
stellar fluxes simultaneously in each image. Unreliable measurements
were culled from the DOLPHOT output using the crowding ($\langle
C_{\rm \lambda} \rangle > 0.25$), sharpness ($\Sigma \vert S_{\rm
  \lambda} \vert >1$), and signal-to-noise ($S/N<3$) parameters,
resulting in a total of $42,940$ sources detected in all four
filters.

We performed false-star tests to determine the photometric completeness
and uncertainty, wherein we added $\approx$50,000 stars (one at a time)
using the appropriate PSFs into each image and reran the photometry as
described above. The fake stars were randomly distributed spatially and
mimic the observed color-magnitude distribution. The resulting 50\%
completeness, or limiting magnitude, is $21.6 < m_{50\%} < 22.8$~mag,
with the bluer filters showing better completeness.

To these data, we match the stellar photometric
catalogs\footnote{http://archive.stsci.edu/prepds/phat/} in the
WFC3/IR F110W and F160W filters from the Panchromatic Hubble Andromeda
Treasury (PHAT) program \citep{PHATpaper}. We made no correction for
interstellar extinction because it is expected to be low in these
filters and will have minimal effect on the separation of C and M
stars in F127M$-$F139M vs. F139M$-$F153M
(Sect.~\ref{sec:cstars}). \citet{Barmby+00} find
  $E(B-V)$ as high as 0.4~mag \citep[or $E(J-K_{\rm s})\lesssim
    0.1$~mag;][]{RiekeLebofsky85} for globular clusters near the observed
  field, which can cause a shift of $<$0.02~mag in these colors.
Figure~\ref{fig:cmds} shows the color-magnitude diagrams (CMDs) and
includes the C star candidates from \citet{Davidge+05} with
$H-K\gtrsim0.4$ \citep[the photometry was redone by][]{Olsen+06} and
candidate C stars identified by the medium-band WFC3/IR colors
(Sect.~\ref{sec:cstars}).  The thin red giant branch indicates minimal
differential extinction across the field.

\subsection{Identifying Carbon Star Candidates}
\label{sec:cstars}

The location of the change in slope in the luminosity function
indicates that the tip of the red giant branch (TRGB) is $M_{\rm
  F153M}^{\rm TRGB} = 18.45$$\pm$$0.10$~mag, or $M_{\rm F153M} =
-6.0$~mag for a distance modulus of 24.45~mag \citep{PHATpaper}. The
vast majority of thermally-pulsing (TP-)AGB stars should lie above
this limit, with exceptions for a subset of stars experiencing minimum
flux during a thermal pulse cycle and very heavily enshrouded
stars. We find 3032 AGB stars that are brighter than
  the TRGB; those that are not classified as C stars are considered to
  be M stars. 

\begin{figure}
\vbox{
\includegraphics[width=\columnwidth]{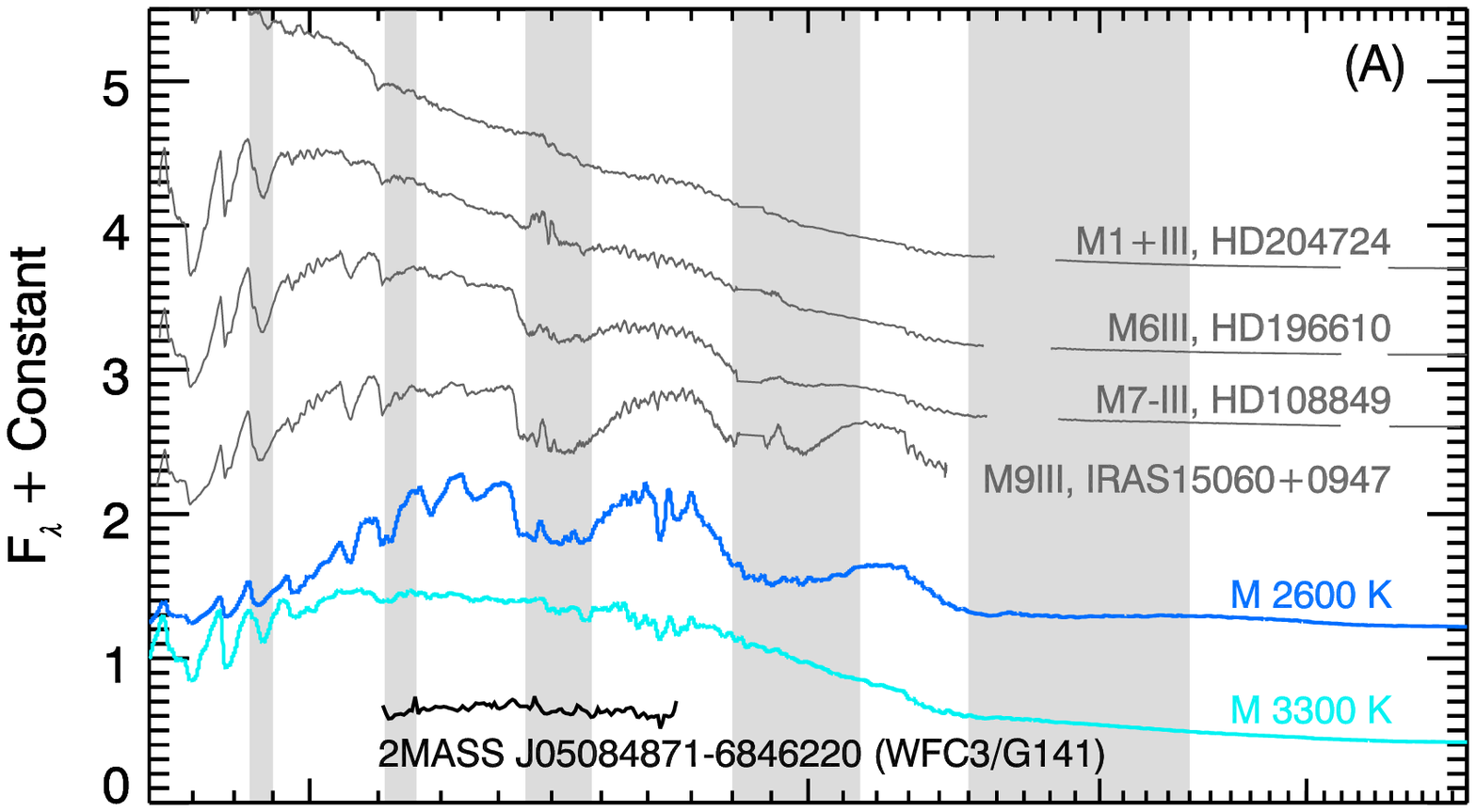}
\includegraphics[width=\columnwidth]{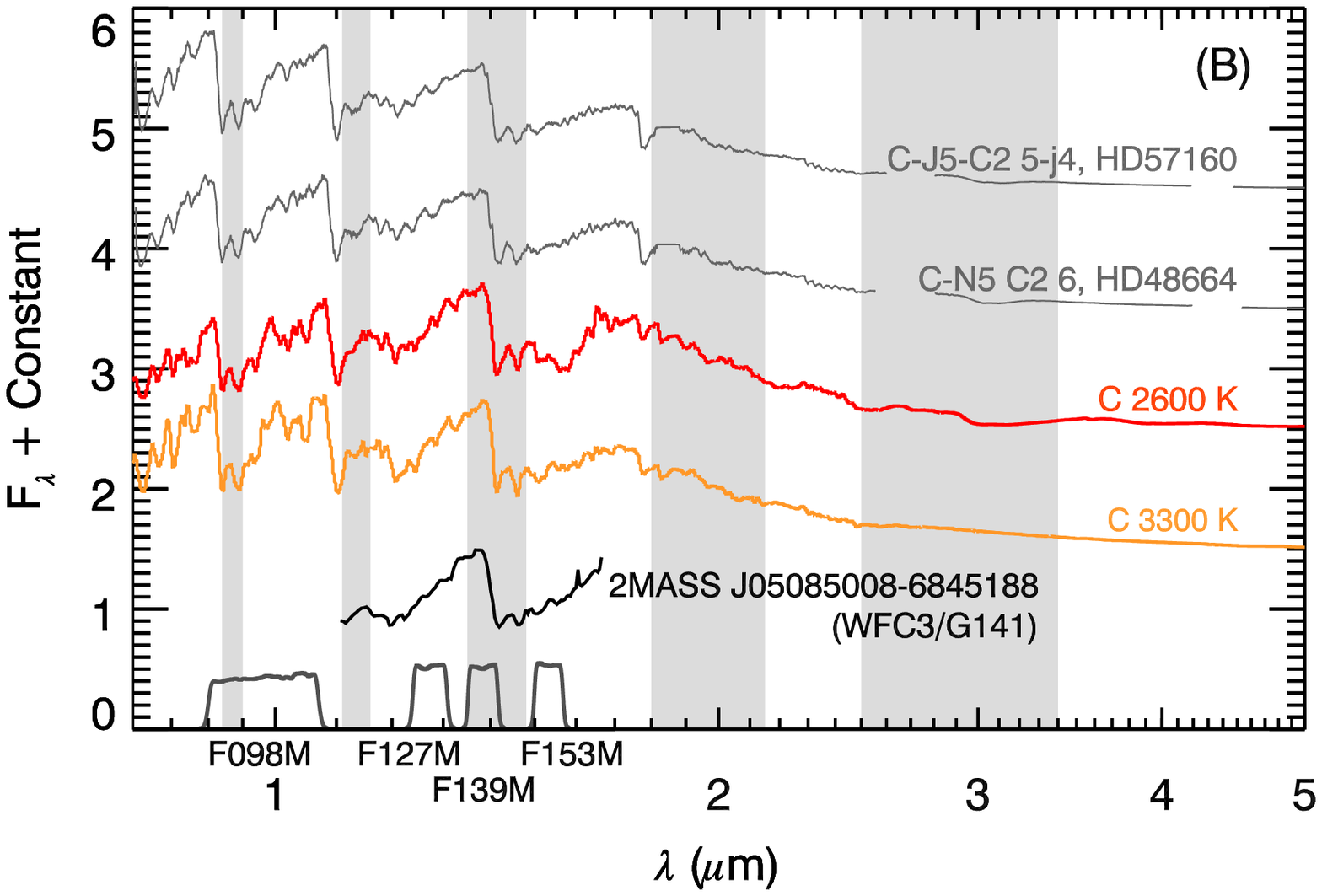}
}
\caption{IRTF spectra of Galactic stars \citep{Rayner+09}, plotted in
  gray, for {\bf (A)} M stars in order of increasing water absorption
  and {\bf (B)} C stars. Solar metallicity model spectra from
  \citet[][and in preparation]{Aringer+09} are also plotted for {\it
    (A)} M stars (blue/cyan) and {\it (B)} C stars (red/orange) with
  $T_{\rm eff} = 2600$~K and $3300$~K. Shaded regions mark wavelengths
  affected by telluric absorption. The model spectra predict all major
  features present in the IRTF spectra. Also shown in black are
  telluric-free spectra of 2 stars in NGC\,1850 across the spectral
  region covered by the WFC3/IR G141 grism: panel {\it (A)} shows the
  M star 2MASS\,J05084871--6846220 and panel {\it (B)} shows the C
  star 2MASS\,J05085008--6845188. These two stars present 2MASS colors
  and magnitudes typical of most C-rich and O-rich AGB stars in the
  LMC and do not show any surprising features in the F139M band.}
\label{fig:IRTF}
\end{figure}

To identify C stars, we employ the color combination of F127M$-$F139M
vs. F139M$-$F153M, which is insensitive to circumstellar
  and interstellar extinction and takes advantage of molecular
features present in AGB spectra. For example, the F153M filter falls
entirely within a broad absorption feature of C$_2$+CN
(Fig.~\ref{fig:IRTF}), with its bandhead at 1.4~\um, which is common
in C-rich stars. O-rich AGB stars of early M subtypes have no
significant absorption feature in the range from $\sim$1 to 1.6~\um,
whereas those of late M subtypes develop a wide H$_2$O absorption
feature between 1.3 and 1.55~\um\ which can be sampled with the F139M
filter. Figure~\ref{fig:ccdmodel}b shows the location of model
atmospheres for objects of M and C type from \citet[][and in
  preparation]{Aringer+09} in color space, as compared to our WFC3
data. Surprisingly, only 1 star resides firmly in the region where C
stars are expected, and only 6 additional sources fall on the edge of
the color-color space covered by the C star models
(Fig.~\ref{fig:ccdmodel}a), far fewer than what is predicted by the
relationship shown in Figure~\ref{fig:cm} (Sect.~\ref{sec:discu}). It
is unclear whether or not the 6 stars near the border drawn in
Fig.~\ref{fig:ccdmodel}b are true carbon stars. However, we verify via
visual inspection that they are real point-sources and are unaffected
by imaging artifacts and/or crowding. To push any of
  these marginal candidates outside of the shaded region via
  correction for interstellar extinction, $E(J-K_{\rm s}) > 0.8$~mag
  is required. This is higher than the expected extinction
  (Sect.~\ref{sec:data}); however, we note that the shaded region in
  Figure~\ref{fig:ccdmodel} is approximate.

To the left of the shaded region in Figure~\ref{fig:ccdmodel} and below
the bulk of the M star population are an additional 8 sources ($0.06
<$ F127M$-$F139M $< 0.18$~mag and $0.13<$ F139M$-$F153M
$<0.23$). Models from Aringer et al.\ (in prep) indicate that these
stars may be M dwarfs.  TRILEGAL simulations in \citet{PHATpaper}
predict that a single WFC3/IR field should contain $\approx$6
foreground M dwarf stars. Of these 8 stars, 5 are covered by the
currently available PHAT catalog, and 3 of these (including the one
closest to the C star region, marked by a blue plus in
Fig.~\ref{fig:ccdmodel}a) have F110W$-$F160W colors consistent with M
dwarf stars \citep[Fig.~\ref{fig:cmds}, also see Fig. 21
  from][]{PHATpaper}. Correction for any possible
  interstellar or circumstellar extinction cannot move these stars
  into the C-star region. We thus conclude that none of these 8 stars
are candidate C stars. We also note that one of the marginal candidate
C stars within the shaded region of Figure~\ref{fig:ccdmodel} shows
F110W$-$F160W$ = 0.57$~mag, a color that is consistent with M dwarf
stars.

Very dusty carbon stars can be undetected even at NIR
  wavelengths.  Reddening vectors in Figure~\ref{fig:ccdmodel} show
  that our method is very insensitive to circumstellar extinction. A
  dust-enshrouded star with $E(J-K_{\rm s}) =1$~mag -- for which any
  color-based C/M classification using ground-based $JHK$ photometry
  would be unreliable -- is marginally affected in the WFC3/IR
  medium-band filters, with color excesses $E({\rm F127M}-{\rm
    F139M})$ and $E({\rm F139M}-{\rm F153M})$ smaller than
  0.2~mag. This has an effect only at the edge of the C star region,
  with the possibility that the 6 stars near the C star
  border are instead very dusty M stars.

In Figure~\ref{fig:ccdmodel}, we show only stars above the
TRGB. However, inclusion of stars up to 1~mag fainter than the F153M
TRGB reveals no additional C star candidates in the shaded region.
Therefore, according to the radiative transfer models of
\citet{Groenewegen06b} and assuming a dust mixture of 70\% amorphous
carbon $+$ 15\% SiC, C stars that are undetected here due to
circumstellar extinction (those fainter than 1~mag below the F153M
TRGB) must have mass-loss rates in excess of
$\sim$$10^{-5}~M_\odot/{\rm yr}$. In the Magellanic Clouds, $<$1.5\%
of C stars have mass-loss rates higher than this \citep[assuming solar
  gas-to-dust ratio, $\psi_\odot = 500$;][]{Boyer+12,Riebel+12}. Thus,
with 1--7 detected C star candidates, we expect $<$0.1 additional
dusty C stars in this field. Including this in the final
  C/M ratio results in ${\rm C/M} = $(1--7.1)$/$(3025--3032)$=
  $($3.3^{+20}_{-0.1}$)$\times$$10^{-4}$, with the lower limit derived
  from the uncertainty in the number of M stars.

\section{Interpretation/discussion}
\label{sec:discu}


A few searches for carbon stars in M31 using optical photometry with
broad and narrow filters have successfully identified AGB
stars. \citet{Brewer+95,Brewer+96} carried out a photometric
narrow-band survey along with a spectroscopic follow up of five fields
southwest of the center of M31 along the major axis
(Fig.~\ref{fig:img}) and found a total of 243 C stars. More recently,
\citet{Nowotny+01}, \citet{Battinelli+03}, and
\citet{BattinelliDemers05} conducted additional searches along the
same southwest axis, using similar techniques. The resulting
relationship between metallicity and the C/M ratio in M31 is shown in
Figure~\ref{fig:cm}.

We can use this relationship to predict the number of C stars expected
in this single field (not including the addition of very dusty,
optically obscured C stars that are missing from
Fig.~\ref{fig:cm}). We assume the surveyed field follows the
metallicity gradient across the bulge, derived by \citet{Saglia+10},
who measure the abundances via optical long-slit data along 6 position
angles, including one that passes very near this field
($0.05\lesssim{\rm [M/H]}\lesssim0.2$ at $R_{\rm M31} = 2$~kpc). A
line fit through all data points in Figure~\ref{fig:cm} thus
predicts 40--64 C stars. The range reflects both the 3\,$\sigma$
uncertainty in the line fit and the range in metallicity.  Using only
the points from \citet{Brewer+95} predicts 11--40 C stars, and the
lowest C/M ratio shown in Figure~\ref{fig:cm} yields 57$\pm$3 C
stars. Here, we investigate the discrepancy between these predictions
and the observed number of C star candidates
(Sect.~\ref{sec:cstars}).



\subsection{Are the missing M31 C stars fainter than our detection limit?}
\label{sec:faint}

Crowding is smooth across the field of view, at approximately 3.5
stars/arcsec$^2$, so it does not affect the completeness at different
positions. Based on false star tests
  (Sect.~\ref{sec:data}), the 50\% completeness limit is
  21.6--22.8~mag, depending on the filter.  Since the TRGB is $>$3~mag
  brighter than the 50\% completeness limit at all positions,
  photometric incompleteness has not affected our C star
  sample. Indeed, the false star tests indicate that the F153M
  photometry is 99\% complete at the TRGB.

Circumstellar extinction could also cause carbon stars to fall below
our detection limit. However, as described in
Section~\ref{sec:cstars}, the extreme reddening required indicates
that we have missed $<$1 C star due to circumstellar dust.



\begin{figure}
\includegraphics[width=\columnwidth]{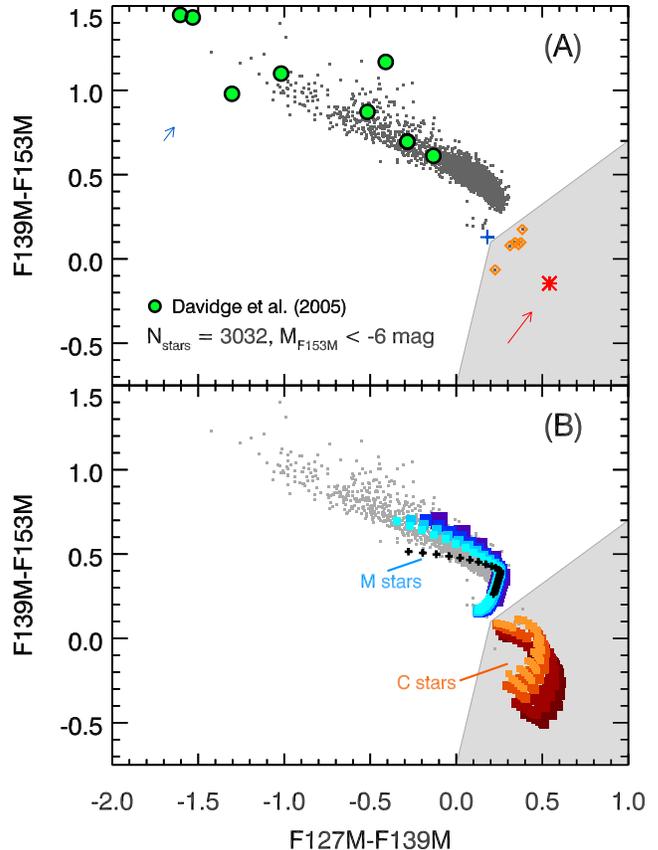}
\caption{Color-color diagram comparing the WFC3/IR colors to the model
  predictions for C and M giants.  {\bf Panel (A)} shows the observed
  colors of AGB stars, with red stars from \citet{Davidge+05} plotted
  as green circles and candidate C stars marked as in
  Figure~\ref{fig:cmds}. Only stars brighter than the TRGB ($M_{\rm
    F153M} < -6$~mag) are included to eliminate most red giant branch
  stars. Circumstellar reddening vectors for
    dust-enshrouded stars are plotted with blue (M star) and red (C
    star) arrows for $E(J-K_{\rm s})= 1.0$~mag (corresponding to a
    mass-loss rate of $\approx$$10^{-6}~M_\odot/{\rm yr}$). These
    vectors are derived from \citet{Groenewegen06b} radiative transfer
    models applied to cool M and C stellar spectra, with 60\% silicate
    $+$ 40\% AlOx, and 70\% amorphous carbon $+$ 15\% SiC mixtures,
    respectively \citep[see][]{Marigo+08}. In {\bf panel (B)}, AGB
  star models are plotted over the data. The M star models (Aringer et
  al., in preparation) are plotted as blue boxes \citep[using the water line list from][]{SCAN}, with darker
  colors representing lower surface gravity, in the range $2 \geq {\rm
    log}\,g~{\rm [cm/s^2]} \geq -0.5$. The \Teff\ decreases from bottom to
  top along this sequence, with values in the range $5000 \leq \Teff
  \leq 2600$~K: the kink with the reddest F127$-$F139M colors is at
  $\Teff\simeq3300$~K. For reference, black pluses show
    an example of the same M star models ($\log g = 0.5$~${\rm
      [cm/s^2]}$) using a different water line list \citep{Barber+06}.
  The C star models \citep{Aringer+09} are plotted as red boxes, with
  darker colors representing higher atmospheric C/O ratios, over the
  range $1.05<{\rm C/O}<2$. They span $4000 \leq
    \Teff\ \leq 2400$~K and $0 \geq \log\,g~{\rm [cm/s^2]} \geq-1$, with
  the coolest C star models now being located at the bottom of the
  drawn sequences. All models assume a solar abundance for the heavy
  metals. The shaded region marks the area where C stars are
  expected.}
\label{fig:ccdmodel}
\end{figure}

\subsection{Are we confident that C-star models are correct?}

Our C-M classification (Fig.~\ref{fig:ccdmodel}) is heavily based on
the behavior of the C and M spectral models from \citet[][and in
  preparation]{Aringer+09}, which should therefore be verified. The
Infrared Telescope Facility (IRTF) database includes NIR spectra of
$\approx$200 Galactic cool giant stars
\citep{Rayner+09}. Figure~\ref{fig:IRTF} illustrates that all observed
features are predicted by the model spectra from \citet[][and in
  preparation]{Aringer+09}. However, because of significant telluric
absorption from water and methane from 1.35--1.42~\um\ ($<$20\%
transmission), spectral features that cause C and M stars to separate
in Figure~\ref{fig:ccdmodel} are not observable from the ground.

As an alternative, we can turn to space-based NIR spectral
observations. Specifically, AGB stars in the LMC cluster NGC\,1850
were observed by the HST program 11913 (PI: J.~MacKenty) with the
WFC3/IR grism G141, thus providing spectra that avoid the problem of
telluric absorption. Figure~\ref{fig:IRTF}b shows the spectra
extracted for the C star 2MASS\,J05085008-6845188
($J-K_\mathrm{s}=1.65$), which reveals no unexpected features in the
F139M filter that might account for the apparent lack of C stars in
Figure~\ref{fig:ccdmodel}.


Bright M stars in the same LMC dataset present
$J-K_\mathrm{s}\simeq1.0$ and generally flat spectra across G141, as
exemplified by the spectrum of 2MASS\,J05084871-6846220 in
Figure~\ref{fig:IRTF}a. These flat spectra are identified with
M-giant models of $\Teff\gtrsim3300$~K, or equivalently with
early-M subtypes, in which the 1.3--1.55~\um\ water absorption feature
is weak.


\subsection{C stars or late M-giants?}

 Our observations show that stars with the reddest $JHK$ colors from
 \citet{Davidge+05} are likely late-M giants (i.e., very cool AGB
 stars, but with O-dominant atmospheres) -- their colors in
 Figure~\ref{fig:ccdmodel} point to the presence of deep water
 absorption features, which depress the F139M flux and cause red
 F139M$-$F153M and blue F127M$-$F139M colors. The M giant sequence is
 well described by Aringer et al.\ (in prep) models using the
 \citet{SCAN} water line list. However, difficulty in modeling
 dynamical atmospheres (pulsation and mass loss) truncates the model
 sequence at F127M$-$${\rm F139M} \approx -0.5$~mag.

\begin{figure}
\includegraphics[width=\columnwidth]{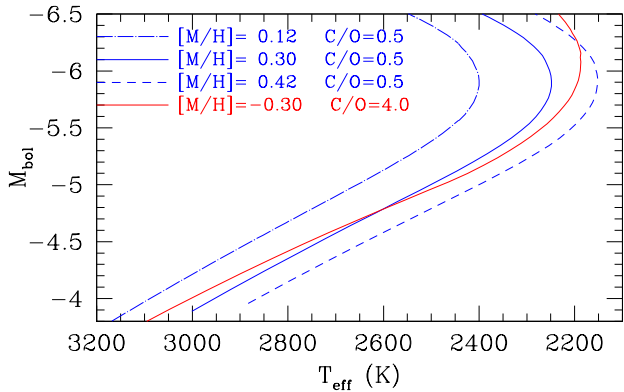}
\caption{TP-AGB Hayashi lines for a model with $M=1.4\, M_{\odot}$ at
  varying initial metallicity ([M/H]) and surface C/O ratio, as
  indicated.  At a given bolometric magnitude, the effective
  temperature is obtained from integrations of complete atmosphere and
  envelope models \citep{Marigo+13} that include on-the-fly
  computation of both molecular chemistry and opacity
  \citep{MarigoAringer09}.
\label{fig:hayashi}}
\end{figure}

 These stars were identified as C star candidates by
 \citet{Davidge+05} since their red colors ($J-K>1.3$~mag) are usually
 markers for C-rich stars in the Magellanic Clouds and other nearby
 galaxies \citep[e.g.,][]{Groenewegen+09,NikolaevWeinberg00}.
 However, our observations clearly indicate that this is not the case
 in M31. In the metal-poor LMC, the O-rich AGB stars are relatively
 warm and rarely reach spectral subtypes later than M3
 \citep[e.g.,][]{Elias+85,Frogel+90,MasseyOlsen03}. On the other hand,
 in the metal rich M31, O-rich AGB stars fall along a cooler Hayashi
 line \citep[cf.][]{Marigo+13}, which ensures a large fraction of them
 appearing as late-M subtypes. The same occurs in our own Galaxy and
 its Bulge, which are known to contain M7--M9 giants in copious
 numbers \citep[e.g.,][]{FrogelWhitford87,Rayner+09}. This fact is
 well represented in Figure~\ref{fig:hayashi}, where we see that
 comparably low effective temperatures of a C-rich model with
 metallicity $[{\rm M/H}] = -0.3$ (typical of the LMC), are reached by
 O-rich models with high metallicity (solar and super-solar,
 reasonably suitable for the inner M31 disk).



\begin{figure}
\includegraphics[width=1\columnwidth]{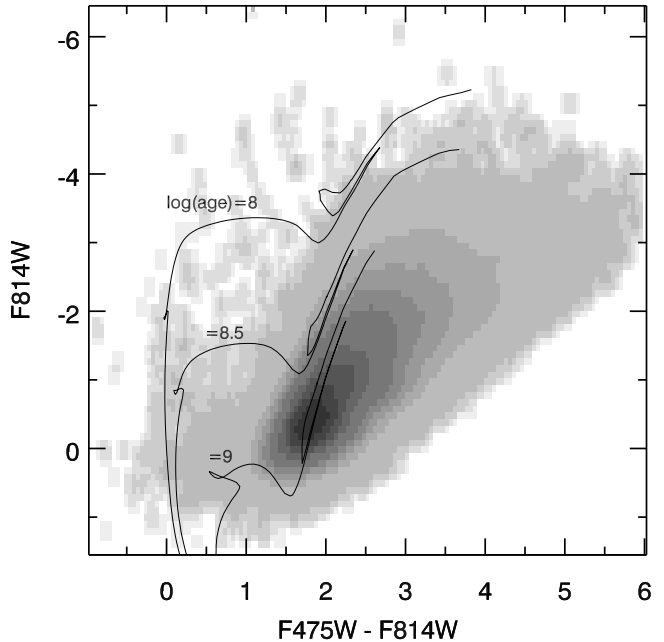}
\caption{Optical CMD of the observed region, from the
    PHAT program \citep{PHATpaper}. Padova isochrones are plotted with
    $A(V) = 0.3$~mag for $\log{\rm(age)} =$ 8, 8.5, and 9
    \citep[PARSEC, version 1.1;][]{Marigo+08,Girardi+10}. The CMD
    indicates the presence of main sequence turn-off stars at least as
    young as $\sim$300~Myr (or $M\lesssim 4~M_\odot$). Detailed
    analysis of this CMD is underway (Williams et al. in prep) to
    quantify the young stellar populations in this region, which is
    not straightforward given the level of crowding in the optical HST
    images.}
\label{fig:opt}
\end{figure}


There are several examples of known late-M giant
  stars that show very red $J-K$ colors
  \citep[e.g.,][]{Marshall+04,vanLoon+05,Javadi+13}, including
  WOH\,G64, an OH/IR star with $T_{\rm eff} \approx 3000$~K. However,
  these red M giants are vastly outnumbered by red C stars in galaxies
  such as the Magellanic Clouds and M33, where most O-rich AGB stars
  are warmer and thus have early M-type spectral classifications (or
  even late-K). Inspection of optical spectra from \citet{Olsen+11} of
  AGB stars in the Magellanic Clouds indicate that only $\approx$3\%
  of M stars show $H-K>0.4$~mag. The HST data indicate
    that this fraction increases to $\approx$30\% in this region of
    M31 (presumably owing to the higher metallicity), assuming that
  stars that show WFC3/IR colors similar to the \citet{Davidge+05}
  stars will also show similar $JHK$ colors.

\subsection{Where are the C stars in this region of M31?}

\paragraph{Age effect?} 
In old, metal-rich stellar populations like the M31
bulge, carbon stars are not expected since the efficiency of the 3DU
($\lambda$) is not sufficient for stars with $M\lesssim2$~M$_\odot$
\citep{Karakas+02}. However, from \citet{SPLASHpaper}, we estimate the
bulge contributes only 28\% of the $I$-band luminosity in this
region. The $I$-band shows minimal contamination from
  short-lived bright sources \citep{MelbourneBoyer13}, so this
  fraction indicates that the bulk of the stellar mass in our observed
  field originates from the younger stellar population of the
  disk. Models from \citet{Courteau+11} fit to the
  3.6~\micron\ luminosity profile also indicate that the disk
  population dominates at 2~kpc.

PHAT optical CMDs for this region (Fig.~\ref{fig:opt}) also reveal the
presence of main sequence turn-off stars younger than $\sim$1~Gyr,
indicating the presence of AGB stars with masses
$\approx$3--5$~M_\odot$.  With a stellar population of this age, we
expect a much larger population of C stars based on the C/M
relationship implied in Figure~\ref{fig:cm}.

\paragraph{Metallicity effect?} 
\citet{Brewer+96} indicate that the C/M ratio rapidly decreases
towards the inner M31 disk (down to values of $\sim$0.019 at $R_{\rm
  M31} = 4$~kpc) and interpret this decrease as being driven by the
metallicity gradient in M31. Our data at $\sim$2~kpc puts additional
constraints in the way the C/M ratio decreases towards the M31
center. Within 2~kpc of M31's center, \citet{Saglia+10} find that the
metallicity spans $0\lesssim {\rm [M/H]} \lesssim 0.5$, with [M/H]
$\approx 0.1 \pm 0.1$ near the position of our observations (see their
Fig.\ 12). If our very low C/M ratio is caused by the high metallicity
at this radius, it supports the presence of a metallicity threshold
above which C stars simply do not form.

\begin{figure}
\vbox{
\includegraphics[width=\columnwidth]{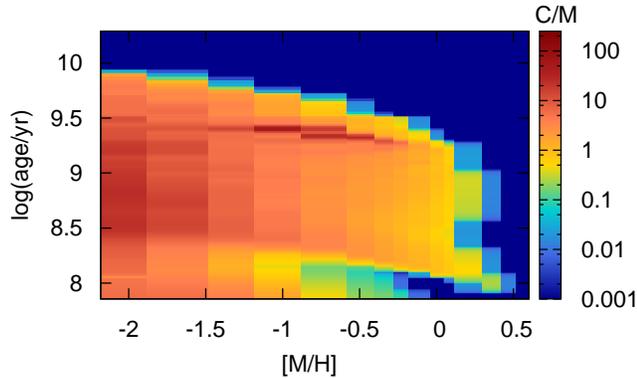}
}
\caption{Map of the C/M ratio in the age-metallicity
    plane, according to the new TP-AGB models computed by Marigo et
    al. (2013).  The lack of the C-star phase (${\rm C/M}=0$)
    corresponds to the darkest blue region, which extends towards
    younger ages at increasing metallicity.  Above some critical value
    of [M/H], models predict C/M$=0$ over the entire age
    range. According to \citet{Saglia+10}, the metallicity of the
    region observed here is $[{\rm M/H}] \approx 0.1 \pm 0.1$.
\label{fig:marigo}}
\end{figure}

A reduction in the C/M ratio with metallicity is indeed indicated by
all modern TP-AGB models \citep{Karakas+02,Marigo+13} as a result of
two effects: the larger amount of carbon that needs to be dredged-up
to make the C/O$>$1 transition, and the fact that the 3DU starts later
(at higher luminosities) and is less efficient at increasing
metallicity.  Figure~\ref{fig:marigo} shows the C/M ratio according to
the new TP-AGB models from \citet{Marigo+13}. For each combination of
age (or initial mass) and initial metallicity, the quantity C/M is
obtained as $\tau_{\rm C}/\tau_{\rm M}$, i.e.  the ratio of the
lifetime spent by the star in the C-rich mode (C/O$>1$) over the one
spent in the O-rich mode (C/O$<1$).  To avoid contamination from the
RGB phase, both $\tau_{\rm C}$ and $\tau_{\rm M}$ are evaluated for
bolometric magnitudes $M_{\rm bol} < -3.6$~mag.  The region of C/M = 0
(dark blue in Fig.~\ref{fig:marigo}) extends to younger ages at
increasing metallicity, encompassing the whole age range for ${\rm
  [M/H]} \gtrsim 0.4$. At these metallicities, more massive TP-AGB
stars (ages $\lesssim 1$~Gyr) go through a few dredge-up events, but
the termination of the TP-AGB phase (determined by the onset of the
super-wind mass loss) occurs before C/O can exceed unity. The models
from \citet{Marigo+13} also indicate that, while the trend depicted in
Figure~\ref{fig:marigo} has a general validity, the exact upper limit
in metallicity for the formation of C stars is sensitive to model
details, mainly related to the 3DU and mass loss. In this context,
observations like the ones we present in this work are of paramount
importance to set stringent constraints on the evolution of TP-AGB
stars in the still largely unexplored regime of high metallicity.

%

\section{Conclusions}
\label{sec:conclu}

We have conducted a search for carbon-rich AGB stars in the inner disk
of M31 using the WFC3/IR medium-band filters on board HST. We find that:

1- WFC3/IR medium filters are very efficient in classifying AGB stars
into their O- and C-rich subtypes, which is important for
extending C star searches in galaxies outside of the Local Group where
resolution and sensitivity prevent ground-based NIR imaging.

2- These filters also place O-rich M stars along a sequence of
increasing water absorption, with the potential of providing
information about the \Teff\ distribution of these stars. Given the
proximity between the F127M and F153M filters, this classification is
largely unaffected by star-to-star variations in the reddening.

3- Contrary to previous work, our observations reveal
the nearly-absence of C stars in this inner disk region of M31. With a
sample of 3032 TP-AGB stars, we find ${\rm (C/M)} =
(3.3^{+20}_{-0.1})$$\times$$10^{-4}$, 1--2 orders of magnitude lower
than previous C star searches in M31.
 
4- This study is among the first to search for complete, unbiased
populations of C stars at ${\rm [M/H]} \gtrsim 0$, and is expected to
provide stringent constraints to evolutionary models of metal-rich AGB
stars. The extremely low observed C/M ratio suggests a metallicity
ceiling for C stars to form. Our results put this ceiling at ${\rm
  [M/H]} \approx 0.1$~dex, which is the estimated mean metallicity of
this M31 region.

5- More observations in these filters across M31 and other galaxies
are strongly needed both for studies of AGB stellar evolution, dust
production, and mass loss.

\acknowledgments

We thank the referee, Jacco van Loon, for thoughtful comments that
improved the manuscript and helped to clarify important issues. This
work was supported by the NASA Postdoctoral Program at the Goddard
Space Flight Center, administered by ORAU through a contract with NASA
and by NASA through HST grant numbers GO-12862 and GO-12055 from the
STScI, which is operated by AURA, Inc., under NASA contract
NAS5-26555. P.M. and L.G. acknowledge support from {\em Progetto di
  Ateneo 2012}, University of Padova, ID: CPDA125588/12. This research
was funded in part by the Austrian Science Fund (FWF): P21988-N16.




{\it Facilities:}  \facility{HST (WFC3)}.






\begin{thebibliography}{65}
\expandafter\ifx\csname natexlab\endcsname\relax\def\natexlab#1{#1}\fi

\bibitem[{{Aringer} {et~al.}(2009){Aringer}, {Girardi}, {Nowotny}, {Marigo}, \&
  {Lederer}}]{Aringer+09}
{Aringer}, B., {Girardi}, L., {Nowotny}, W., {Marigo}, P., \& {Lederer}, M.~T.
  2009, \aap, 503, 913

\bibitem[{{Azzopardi} {et~al.}(1991){Azzopardi}, {Rebeirot}, {Lequeux}, \&
  {Westerlund}}]{Azzopardi+91}
{Azzopardi}, M., {Rebeirot}, E., {Lequeux}, J., \& {Westerlund}, B.~E. 1991,
  \aaps, 88, 265

\bibitem[{{Barber} {et~al.}(2006){Barber}, {Tennyson}, {Harris}, \&
  {Tolchenov}}]{Barber+06}
{Barber}, R.~J., {Tennyson}, J., {Harris}, G.~J., \& {Tolchenov}, R.~N. 2006,
  \mnras, 368, 1087

\bibitem[{{Barmby} {et~al.}(2000){Barmby}, {Huchra}, {Brodie}, {Forbes},
  {Schroder}, \& {Grillmair}}]{Barmby+00}
{Barmby}, P., {Huchra}, J.~P., {Brodie}, J.~P., {Forbes}, D.~A., {Schroder},
  L.~L., \& {Grillmair}, C.~J. 2000, \aj, 119, 727

\bibitem[{{Barmby} {et~al.}(2006){Barmby}, {Ashby}, {Bianchi}, {Engelbracht},
  {Gehrz}, {Gordon}, {Hinz}, {Huchra}, {Humphreys}, {Pahre},
  {P{\'e}rez-Gonz{\'a}lez}, {Polomski}, {Rieke}, {Thilker}, {Willner}, \&
  {Woodward}}]{Barmby+06}
{Barmby}, P., {et~al.} 2006, \apjl, 650, L45

\bibitem[{{Battinelli} \& {Demers}(2005)}]{BattinelliDemers05}
{Battinelli}, P., \& {Demers}, S. 2005, \aap, 430, 905

\bibitem[{{Battinelli} \& {Demers}(2009)}]{BattinelliDemers09}
---. 2009, \aap, 493, 1075

\bibitem[{{Battinelli} {et~al.}(2003){Battinelli}, {Demers}, \&
  {Letarte}}]{Battinelli+03}
{Battinelli}, P., {Demers}, S., \& {Letarte}, B. 2003, \aj, 125, 1298

\bibitem[{{Battinelli} {et~al.}(2007){Battinelli}, {Demers}, \&
  {Mannucci}}]{Battinelli+07}
{Battinelli}, P., {Demers}, S., \& {Mannucci}, F. 2007, \aap, 474, 35

\bibitem[{{Boyer} {et~al.}(2009){Boyer}, {Skillman}, {van Loon}, {Gehrz}, \&
  {Woodward}}]{Boyer+09}
{Boyer}, M.~L., {Skillman}, E.~D., {van Loon}, J.~{\relax Th}., {Gehrz}, R.~D.,
  \& {Woodward}, C.~E. 2009, \apj, 697, 1993

\bibitem[{{Boyer} {et~al.}(2011){Boyer}, {Srinivasan}, {van Loon}, {McDonald},
  {Meixner}, {Zaritsky}, {Gordon}, {Kemper}, {Babler}, {Block}, {Bracker},
  {Engelbracht}, {Hora}, {Indebetouw}, {Meade}, {Misselt}, {Robitaille},
  {Sewi{\l}o}, {Shiao}, \& {Whitney}}]{Boyer+11}
{Boyer}, M.~L., {et~al.} 2011, \aj, 142, 103

\bibitem[{{Boyer} {et~al.}(2012){Boyer}, {Srinivasan}, {Riebel}, {McDonald},
  {van Loon}, {Clayton}, {Gordon}, {Meixner}, {Sargent}, \& {Sloan}}]{Boyer+12}
---. 2012, \apj, 748, 40

\bibitem[{{Brewer} {et~al.}(1995){Brewer}, {Richer}, \& {Crabtree}}]{Brewer+95}
{Brewer}, J.~P., {Richer}, H.~B., \& {Crabtree}, D.~R. 1995, \aj, 109, 2480

\bibitem[{{Brewer} {et~al.}(1996){Brewer}, {Richer}, \& {Crabtree}}]{Brewer+96}
---. 1996, \aj, 112, 491

\bibitem[{{Cioni} {et~al.}(2006){Cioni}, {Girardi}, {Marigo}, \&
  {Habing}}]{Cioni+06}
{Cioni}, M.-R.~L., {Girardi}, L., {Marigo}, P., \& {Habing}, H.~J. 2006, \aap,
  448, 77

\bibitem[{{Conroy} {et~al.}(2009){Conroy}, {Gunn}, \& {White}}]{Conroy+09}
{Conroy}, C., {Gunn}, J.~E., \& {White}, M. 2009, \apj, 699, 486

\bibitem[{{Courteau} {et~al.}(2011){Courteau}, {Widrow}, {McDonald},
  {Guhathakurta}, {Gilbert}, {Zhu}, {Beaton}, \& {Majewski}}]{Courteau+11}
{Courteau}, S., {Widrow}, L.~M., {McDonald}, M., {Guhathakurta}, P., {Gilbert},
  K.~M., {Zhu}, Y., {Beaton}, R.~L., \& {Majewski}, S.~R. 2011, \apj, 739, 20

\bibitem[{{Dalcanton} {et~al.}(2012{\natexlab{a}}){Dalcanton}, {Williams},
  {Melbourne}, {Girardi}, {Dolphin}, {Rosenfield}, {Boyer}, {de Jong},
  {Gilbert}, {Marigo}, {Olsen}, {Seth}, \& {Skillman}}]{SNAPpaper}
{Dalcanton}, J.~J., {et~al.} 2012{\natexlab{a}}, \apjs, 198, 6

\bibitem[{{Dalcanton} {et~al.}(2012{\natexlab{b}}){Dalcanton}, {Williams},
  {Lang}, {Lauer}, {Kalirai}, {Seth}, {Dolphin}, {Rosenfield}, {Weisz}, {Bell},
  {Bianchi}, {Boyer}, {Caldwell}, {Dong}, {Dorman}, {Gilbert}, {Girardi},
  {Gogarten}, {Gordon}, {Guhathakurta}, {Hodge}, {Holtzman}, {Johnson},
  {Larsen}, {Lewis}, {Melbourne}, {Olsen}, {Rix}, {Rosema}, {Saha},
  {Sarajedini}, {Skillman}, \& {Stanek}}]{PHATpaper}
---. 2012{\natexlab{b}}, \apjs, 200, 18

\bibitem[{{Davidge} {et~al.}(2005){Davidge}, {Olsen}, {Blum}, {Stephens}, \&
  {Rigaut}}]{Davidge+05}
{Davidge}, T.~J., {Olsen}, K.~A.~G., {Blum}, R., {Stephens}, A.~W., \&
  {Rigaut}, F. 2005, \aj, 129, 201

\bibitem[{{Dolphin}(2000)}]{Dolphin00}
{Dolphin}, A.~E. 2000, PASP, 112, 1383

\bibitem[{{Dorman} {et~al.}(2012){Dorman}, {Guhathakurta}, {Fardal}, {Lang},
  {Geha}, {Howley}, {Kalirai}, {Bullock}, {Cuillandre}, {Dalcanton}, {Gilbert},
  {Seth}, {Tollerud}, {Williams}, \& {Yniguez}}]{SPLASHpaper}
{Dorman}, C.~E., {et~al.} 2012, \apj, 752, 147

\bibitem[{{Elias} {et~al.}(1985){Elias}, {Frogel}, \& {Humphreys}}]{Elias+85}
{Elias}, J.~H., {Frogel}, J.~A., \& {Humphreys}, R.~M. 1985, \apjs, 57, 91

\bibitem[{{Feast}(2007)}]{Feast07}
{Feast}, M. 2007, in Astronomical Society of the Pacific Conference Series,
  Vol. 378, Why Galaxies Care About AGB Stars: Their Importance as Actors and
  Probes, ed. F.~{Kerschbaum}, C.~{Charbonnel}, \& R.~F. {Wing}, 479

\bibitem[{{Ferrarotti} \& {Gail}(2006)}]{FerrarottiGail06}
{Ferrarotti}, A.~S., \& {Gail}, H.-P. 2006, \aap, 447, 553

\bibitem[{{Frogel} {et~al.}(1990){Frogel}, {Mould}, \& {Blanco}}]{Frogel+90}
{Frogel}, J.~A., {Mould}, J., \& {Blanco}, V.~M. 1990, \apj, 352, 96

\bibitem[{{Frogel} \& {Whitford}(1987)}]{FrogelWhitford87}
{Frogel}, J.~A., \& {Whitford}, A.~E. 1987, \apj, 320, 199

\bibitem[{{Girardi} {et~al.}(2010){Girardi}, {Williams}, {Gilbert},
  {Rosenfield}, {Dalcanton}, {Marigo}, {Boyer}, {Dolphin}, {Weisz},
  {Melbourne}, {Olsen}, {Seth}, \& {Skillman}}]{Girardi+10}
{Girardi}, L., {et~al.} 2010, \apj, 724, 1030

\bibitem[{{Groenewegen}(2006{\natexlab{a}})}]{Groenewegen06}
{Groenewegen}, M.~A.~T. 2006{\natexlab{a}}, in Planetary Nebulae Beyond the
  Milky Way, ed. L.~{Stanghellini}, J.~R. {Walsh}, \& N.~G. {Douglas}, 108

\bibitem[{{Groenewegen}(2006{\natexlab{b}})}]{Groenewegen06b}
{Groenewegen}, M.~A.~T. 2006{\natexlab{b}}, \aap, 448, 181

\bibitem[{{Groenewegen}(2007)}]{Groenewegen07}
{Groenewegen}, M.~A.~T. 2007, in Astronomical Society of the Pacific Conference
  Series, Vol. 378, Why Galaxies Care About AGB Stars: Their Importance as
  Actors and Probes, ed. F.~{Kerschbaum}, C.~{Charbonnel}, \& R.~F. {Wing}, 433

\bibitem[{{Groenewegen} {et~al.}(2009){Groenewegen}, {Sloan}, {Soszy{\'n}ski},
  \& {Petersen}}]{Groenewegen+09}
{Groenewegen}, M.~A.~T., {Sloan}, G.~C., {Soszy{\'n}ski}, I., \& {Petersen},
  E.~A. 2009, \aap, 506, 1277

\bibitem[{{Hook} {et~al.}(2008){Hook}, {Stoehr}, \& {Krist}}]{Hook+08}
{Hook}, R., {Stoehr}, F., \& {Krist}, J. 2008, Space Telescope European
  Coordinating Facility Newsletter, 44, 11

\bibitem[{{Javadi} {et~al.}(2013){Javadi}, {van Loon}, {Khosroshahi}, \&
  {Mirtorabi}}]{Javadi+13}
{Javadi}, A., {van Loon}, J.~{\relax Th}., {Khosroshahi}, H., \& {Mirtorabi},
  M.~T. 2013, \mnras, 432, 2824

\bibitem[{{J{\o}rgensen} {et~al.}(2001){J{\o}rgensen}, {Jensen}, {S{\o}rensen},
  \& {Aringer}}]{SCAN}
{J{\o}rgensen}, U.~G., {Jensen}, P., {S{\o}rensen}, G.~O., \& {Aringer}, B.
  2001, \aap, 372, 249

\bibitem[{{Karakas} \& {Lattanzio}(2007)}]{KarakasLattanzio07}
{Karakas}, A.~I., \& {Lattanzio}, J.~C. 2007, PASA, 24, 103

\bibitem[{{Karakas} {et~al.}(2002){Karakas}, {Lattanzio}, \&
  {Pols}}]{Karakas+02}
{Karakas}, A.~I., {Lattanzio}, J.~C., \& {Pols}, O.~R. 2002, PASA, 19, 515

\bibitem[{{Krist}(1995)}]{Krist95}
{Krist}, J. 1995, in Astronomical Society of the Pacific Conference Series,
  Vol.~77, Astronomical Data Analysis Software and Systems IV, ed. R.~A.
  {Shaw}, H.~E. {Payne}, \& J.~J.~E. {Hayes}, 349

\bibitem[{{Maraston} {et~al.}(2006){Maraston}, {Daddi}, {Renzini}, {Cimatti},
  {Dickinson}, {Papovich}, {Pasquali}, \& {Pirzkal}}]{Maraston+06}
{Maraston}, C., {Daddi}, E., {Renzini}, A., {Cimatti}, A., {Dickinson}, M.,
  {Papovich}, C., {Pasquali}, A., \& {Pirzkal}, N. 2006, \apj, 652, 85

\bibitem[{{Marigo}(2001)}]{Marigo01}
{Marigo}, P. 2001, \aap, 370, 194

\bibitem[{{Marigo} \& {Aringer}(2009)}]{MarigoAringer09}
{Marigo}, P., \& {Aringer}, B. 2009, \aap, 508, 1539

\bibitem[{{Marigo} {et~al.}(2013){Marigo}, {Bressan}, {Nanni}, {Girardi}, \&
  {Pumo}}]{Marigo+13}
{Marigo}, P., {Bressan}, A., {Nanni}, A., {Girardi}, L., \& {Pumo}, M.~L. 2013,
  MNRAS, in press, arXiv:astro-ph/1305.4485

\bibitem[{{Marigo} {et~al.}(2008){Marigo}, {Girardi}, {Bressan}, {Groenewegen},
  {Silva}, \& {Granato}}]{Marigo+08}
{Marigo}, P., {Girardi}, L., {Bressan}, A., {Groenewegen}, M.~A.~T., {Silva},
  L., \& {Granato}, G.~L. 2008, \aap, 482, 883

\bibitem[{{Marigo} {et~al.}(2010){Marigo}, {Girardi}, {Bressan}, {Aringer},
  {Gullieuszik}, {Held}, {Groenewegen}, {Silva}, \& {Granato}}]{Marigo+10}
{Marigo}, P., {et~al.} 2010, in IAU Symposium, Vol. 262, IAU Symposium, ed.
  G.~R. {Bruzual} \& S.~{Charlot}, 36--43

\bibitem[{{Marshall} {et~al.}(2004){Marshall}, {van Loon}, {Matsuura}, {Wood},
  {Zijlstra}, \& {Whitelock}}]{Marshall+04}
{Marshall}, J.~R., {van Loon}, J.~{\relax Th}., {Matsuura}, M., {Wood}, P.~R.,
  {Zijlstra}, A.~A., \& {Whitelock}, P.~A. 2004, \mnras, 355, 1348

\bibitem[{{Massey} \& {Olsen}(2003)}]{MasseyOlsen03}
{Massey}, P., \& {Olsen}, K.~A.~G. 2003, \aj, 126, 2867

\bibitem[{{Melbourne} \& {Boyer}(2013)}]{MelbourneBoyer13}
{Melbourne}, J., \& {Boyer}, M.~L. 2013, \apj, 764, 30

\bibitem[{{Melbourne} {et~al.}(2012){Melbourne}, {Williams}, {Dalcanton},
  {Rosenfield}, {Girardi}, {Marigo}, {Weisz}, {Dolphin}, {Boyer}, {Olsen},
  {Skillman}, \& {Seth}}]{Melbourne+12}
{Melbourne}, J., {et~al.} 2012, \apj, 748, 47

\bibitem[{{Miszalski} {et~al.}(2013){Miszalski}, {Miko{\l}ajewska}, \&
  {Udalski}}]{Miszalski+13}
{Miszalski}, B., {Miko{\l}ajewska}, J., \& {Udalski}, A. 2013, \mnras, 432,
  3186

\bibitem[{{Nikolaev} \& {Weinberg}(2000)}]{NikolaevWeinberg00}
{Nikolaev}, S., \& {Weinberg}, M.~D. 2000, \apj, 542, 804

\bibitem[{{Nowotny} {et~al.}(2013){Nowotny}, {Aringer}, {H{\"o}fner}, \&
  {Eriksson}}]{Nowotny+13}
{Nowotny}, W., {Aringer}, B., {H{\"o}fner}, S., \& {Eriksson}, K. 2013, \aap,
  552, A20

\bibitem[{{Nowotny} {et~al.}(2001){Nowotny}, {Kerschbaum}, {Schwarz}, \&
  {Olofsson}}]{Nowotny+01}
{Nowotny}, W., {Kerschbaum}, F., {Schwarz}, H.~E., \& {Olofsson}, H. 2001,
  \aap, 367, 557

\bibitem[{{Olsen} {et~al.}(2006){Olsen}, {Blum}, {Stephens}, {Davidge},
  {Massey}, {Strom}, \& {Rigaut}}]{Olsen+06}
{Olsen}, K.~A.~G., {Blum}, R.~D., {Stephens}, A.~W., {Davidge}, T.~J.,
  {Massey}, P., {Strom}, S.~E., \& {Rigaut}, F. 2006, \aj, 132, 271

\bibitem[{{Olsen} {et~al.}(2011){Olsen}, {Zaritsky}, {Blum}, {Boyer}, \&
  {Gordon}}]{Olsen+11}
{Olsen}, K.~A.~G., {Zaritsky}, D., {Blum}, R.~D., {Boyer}, M.~L., \& {Gordon},
  K.~D. 2011, \apj, 737, 29

\bibitem[{{Rayner} {et~al.}(2009){Rayner}, {Cushing}, \& {Vacca}}]{Rayner+09}
{Rayner}, J.~T., {Cushing}, M.~C., \& {Vacca}, W.~D. 2009, \apjs, 185, 289

\bibitem[{{Riebel} {et~al.}(2012){Riebel}, {Srinivasan}, {Sargent}, \&
  {Meixner}}]{Riebel+12}
{Riebel}, D., {Srinivasan}, S., {Sargent}, B., \& {Meixner}, M. 2012, \apj,
  753, 71

\bibitem[{{Rieke} \& {Lebofsky}(1985)}]{RiekeLebofsky85}
{Rieke}, G.~H., \& {Lebofsky}, M.~J. 1985, \apj, 288, 618

\bibitem[{{Rowe} {et~al.}(2005){Rowe}, {Richer}, {Brewer}, \&
  {Crabtree}}]{Rowe+05}
{Rowe}, J.~F., {Richer}, H.~B., {Brewer}, J.~P., \& {Crabtree}, D.~R. 2005,
  \aj, 129, 729

\bibitem[{{Saglia} {et~al.}(2010){Saglia}, {Fabricius}, {Bender}, {Montalto},
  {Lee}, {Riffeser}, {Seitz}, {Morganti}, {Gerhard}, \& {Hopp}}]{Saglia+10}
{Saglia}, R.~P., {et~al.} 2010, \aap, 509, A61

\bibitem[{{Sibbons} {et~al.}(2012){Sibbons}, {Ryan}, {Cioni}, {Irwin}, \&
  {Napiwotzki}}]{Sibbons+12}
{Sibbons}, L.~F., {Ryan}, S.~G., {Cioni}, M.-R.~L., {Irwin}, M., \&
  {Napiwotzki}, R. 2012, \aap, 540, A135

\bibitem[{{Stephens} {et~al.}(2003){Stephens}, {Frogel}, {DePoy}, {Freedman},
  {Gallart}, {Jablonka}, {Renzini}, {Rich}, \& {Davies}}]{Stephens+03}
{Stephens}, A.~W., {et~al.} 2003, \aj, 125, 2473

\bibitem[{{van Loon} {et~al.}(2005){van Loon}, {Cioni}, {Zijlstra}, \&
  {Loup}}]{vanLoon+05}
{van Loon}, J.~{\relax Th}., {Cioni}, M.-R.~L., {Zijlstra}, A.~A., \& {Loup},
  C. 2005, \aap, 438, 273

\bibitem[{{Ventura} {et~al.}(2001){Ventura}, {D'Antona}, {Mazzitelli}, \&
  {Gratton}}]{Ventura+01}
{Ventura}, P., {D'Antona}, F., {Mazzitelli}, I., \& {Gratton}, R. 2001, \apjl,
  550, L65

\bibitem[{{Whitelock}(1993)}]{Whitelock+93}
{Whitelock}, P. 1993, in IAU Symposium, Vol. 153, Galactic Bulges, ed.
  H.~{Dejonghe} \& H.~J. {Habing}, 39

\bibitem[{{Zurita} \& {Bresolin}(2012)}]{ZuritaBresolin12}
{Zurita}, A., \& {Bresolin}, F. 2012, \mnras, 427, 1463

\end{thebibliography}


\end{document}